\documentclass[mypaper,7pt,twoside]{CoAst}
\usepackage{epsf,graphicx,fancyhdr,natbib}
\renewcommand{\chaptermark}[1]{\markboth{#1}{}}

\newcommand{\kopf}{\small\itshape Comm. in Asteroseismology\\ Vol. 148, 2006}
\newcommand{\Authors}[1]{\begin{center}\normalsize\bf\sf #1 \end{center}}

\renewcommand{\author}[1]{\begin{center}\normalsize\bf\sf #1 \end{center}}
\newcommand{\Address}[1]{\begin{center}\small\sf #1 \end{center}}

\renewenvironment{abstract}{\section*{Abstract}\normalsize\sf}{}
\newcommand{\References}[1]{\begin{flushleft}{\large References\\}\vspace*{2mm}\small #1 \end{flushleft}}

\newcommand{\chapterDSSN}[2]{\chapter[\sf\normalsize #1\\ \footnotesize \hspace*{5mm}by #2 \sf\normalsize][]{#1\\}\rhead[\fancyplain{}{\sf\footnotesize \center{#1}}]{\fancyplain{}{\sffamily\thepage}}\lhead[\fancyplain{\kopf}{\sffamily\thepage}]{\fancyplain{\kopf}{\sf\footnotesize \center{#2}}}}

\newcommand{\figureDSSN}[5]{\begin{figure}[#4]
\centering
\includegraphics*[#5]{#1}
\caption{#2}
\label{#3}
\end{figure}}

\renewcommand{\chaptername}{}
\newcommand{\acknowledgments}[1]{\vspace*{5mm}\noindent\begin{bf}Acknowledgments. \end{bf} #1}

\begin{document}
\sf

\chapterDSSN{Theory of rapidly oscillating Ap stars}{M.S.~Cunha}

\Authors{M.S.~Cunha} 
\Address{Centro de Astrof\'\i sica da Universidade do Porto, Portugal and High Altitude Observatory, Boulder, USA}

\noindent
\begin{abstract}
I review recent theoretical work on rapidly oscillating Ap stars and discuss key aspects of the physics of the oscillations observed in this class of pulsators.
\end{abstract}

\section*{1 Introduction}

Rapidly oscillating Ap stars (hereafter roAp stars) are main-sequence chemically peculiar stars of spectral type A (and sometimes F), which exhibit oscillations with amplitudes of a few mmag and frequencies typically ranging from 1 to 3~mHz. Over 30 stars are presently known to belong to this class of pulsators, the first examples having been discovered almost three decades ago by \cite{kurtz82}. Moreover, lower-frequency pulsators have been predicted  to exist among the more evolved cool Ap stars \citep{cunha02}. A first example of the latter has recently been discovered \citep{elkin05}, showing that the roAp phenomena is likely to span a frequency range which is wider than usually considered. 

In the past few years a number of exciting observational results have motivated the development of theoretical work on roAp stars. In particular, high time- and spectral-resolution spectroscopic observations have unveiled the oscillations in the atmosphere of these stars (Kochukhov, {\it these proceedings}), and high duty cycle photometric data, acquired either with the ground-based telescope network WET \citep{kurtz05}, or with the Canadian satellite MOST (Matthews {\it these proceedings}), have improved the oscillation spectra of particular members of this class of pulsators.  

 Additional recent observational and theoretical reviews on roAp stars are provided in \cite{kurtz04}, \cite{cunha03,cunha05} and \cite{gough05}.

\section*{2 The importance of the magnetic field}

Ap stars are known to have strong, well organized magnetic fields with typical magnitudes of a few kG \citep{mathys97,hubrig04,hubrig05,kochukhov06,ryabchikova06}. Well below the photosphere the magnetic field is unlikely to play an important role in the dynamics of the oscillations. However, in the outer layers it is expected to influence the oscillations both directly, through the action of an additional restoring force \citep[e.g.][]{dziembowski96,cunha00}, and indirectly, through the interaction with outer convection \citep{balmforth01}.  

Figure \ref{pressure} shows the outer layers of a typical roAp star, where the magnetic field is likely to influence the pulsations, for different magnetic field intensities. In particular, two different regions of influence should be considered: the magnetoacoustic layer, where the magnetic and gas pressures are of the same order of magnitude, and, above it, the magnetically dominated layer, where the former becomes much larger than the latter. These two regions together form what we shall designate the {\it magnetic boundary layer}.

\figureDSSN{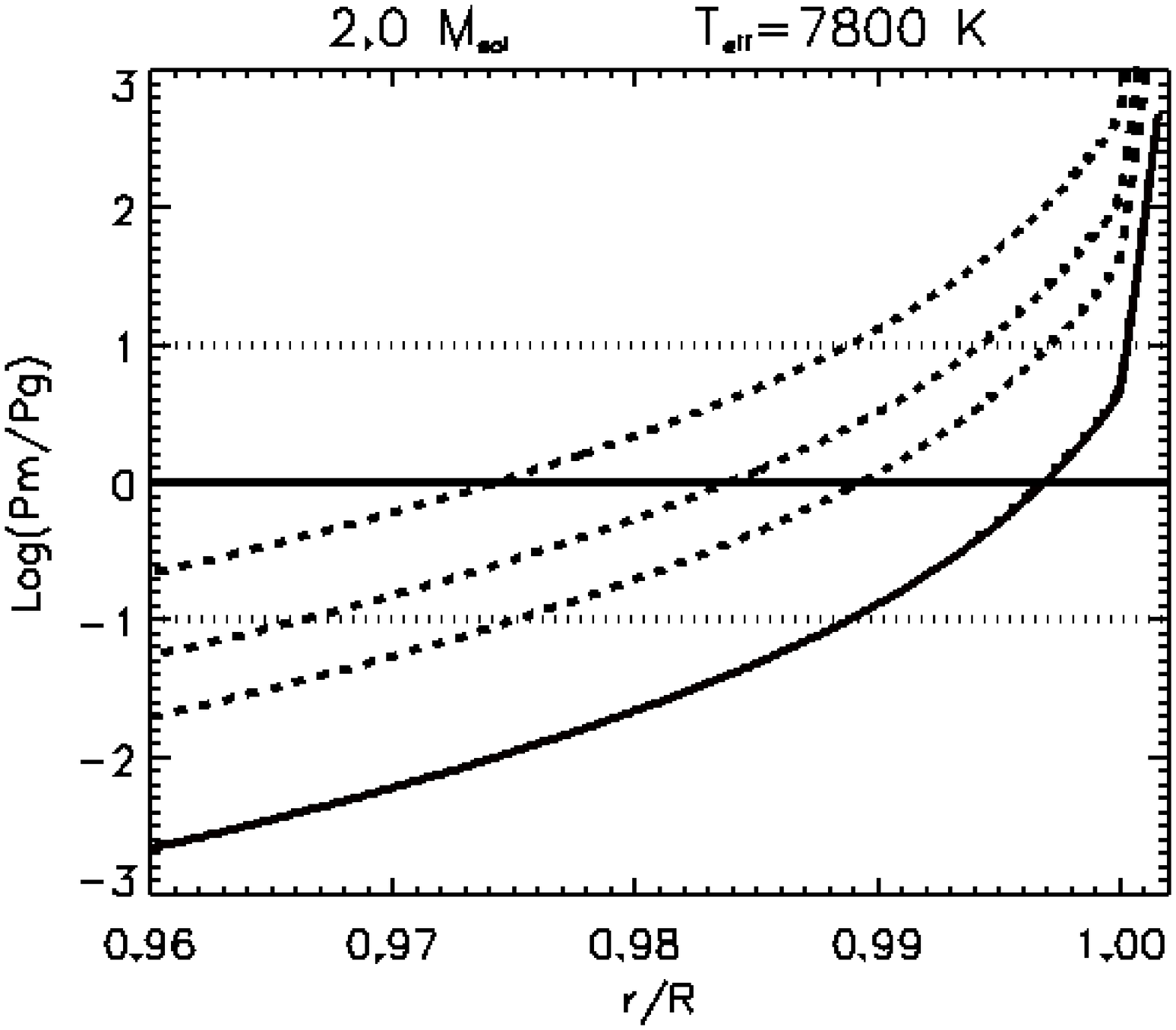}{Ratio between magnetic (Pm) and gas (Pg) pressures for different magnetic field intensities. The two horizontal dashed lines bound the region where the magnetic and acoustic pressure are of the same order of magnitude. The continuous line shows the pressure ratio for a magnetic field strength of 1kG, while the additional dashed lines show the pressure ratio for fields of magnitude 3kG, 5kG and 10kG, respectively.}{pressure}{!ht}{clip,angle=0,width=67mm}

\subsection*{2.1 Direct effect on atmospheric pulsations}

Over the past few years, exciting results have been derived through the analysis of time-series of high resolution spectroscopic data of roAp stars. These data contain information about the structure of pulsations in the atmosphere of these stars. Here we emphasize some aspects that are likely to be important when interpreting these exciting data.

Figure \ref{pressure} shows that the photosphere of a roAp star (indicated by the change in the slope of the curves) can be located either in the magnetoacoustic, or in the magnetically dominated region, depending on the strength of the magnetic field. Thus, depending on the latter, the structure of the oscillations in the atmosphere of these stars might look significantly different. While in the magnetoacoustic region the restoring force has acoustic and magnetic components of comparable size, in the magnetically dominated region the magnetoacoustic wave decouples, and we may expect to find waves which are essentially magnetic and waves which are essentially acoustic. 

In the latter regime, the direction of the displacement associated with each of the components is determined by the direction of the perturbed Lorentz force, which, to first order, is perpendicular to the unperturbed magnetic field. Thus, the acoustic and magnetic components will have associated displacements, respectively, along, and perpendicular to the direction of the magnetic field lines. Unless the structure of the atmosphere is very different from that currently accepted, when observed, the acoustic waves are expected to be in the form of running waves, with frequency larger than the local acoustic cutoff frequency. Since the latter depends on the inclination of the magnetic field \citep[e.g.][]{dziembowski96}, becoming smaller as the inclination of the field in relation to the local vertical increases, one would expect to find these running waves only at particular latitudes, which will depend on the frequency of the oscillation considered. Moreover, since the acoustic displacement is forced to be along the magnetic field lines, its phase, at a given depth and moment in time, is expected to depend on latitude. Hence, when interpreting disk-averaged data it is important to keep in mind that very different displacements might be expected at different latitudes and that aspects such as the surface distribution of the elements are likely to influence significantly what is observed. 

When the magnetic field is sufficiently weak, the photosphere will be located in the magnetoacoustic region and one can no longer describe the displacement there as a simple superposition of magnetic and acoustic components. It is beyond the scope of this paper to analise how the  displacement in this region would be seen in disk-averaged high resolution spectroscopic data. However, it might be enlightening in this context to revisit the study of pulsations in a simple toy model, composed of two adjacent isothermal layers with different characteristic sound speeds \citep[e.g.][]{balmforth90}. If in the top layer the waves are allowed to propagate away, then in the lower layer, which is assumed to have a fully reflected lower boundary, the displacement can be expressed as a sum of a standing wave and a running wave. No matter how small the running wave component might be, if the sanding wave component goes through a node at a given depth, at that depth the running wave component will dominate the solution for the displacement. Hence, if one could 'observe' the waves in the lower layer of this toy model and match the observations to a function of the form $A$cos($\omega t+\phi)$, with $\omega$ being the frequency of the oscillation, $t$ the time and $\phi$ a phase, one would find $\phi$ to be almost constant at all depths, except close to the nodes of the standing component, where the latter would change significantly with depth. Naturally, in the magnetoacoustic layer of roAp stars there are additional elements that need to be kept in mind. In particular, just as in the case of the magnetically dominated region, the depth structure of the displacement is expected to depend on latitude, and any analysis of disk-average data aiming at reconstructing the form of the eigenfunctions in this region must take that dependence into consideration.

\subsection*{2.2 Direct effect on the global properties of pulsations}

The magnetic boundary layer influences basic properties of the pulsations such as the oscillation frequencies and eigenfunctions. These effects cannot be neglected when attempting to use common asteroseismic tools, such as large and small separations, to infer information about the properties of the interior of these stars. On the other hand, these magnetic signatures contain information about properties of the magnetic field, which might be extracted if only we understand the way in which the magnetic field influences the pulsations.

\figureDSSN{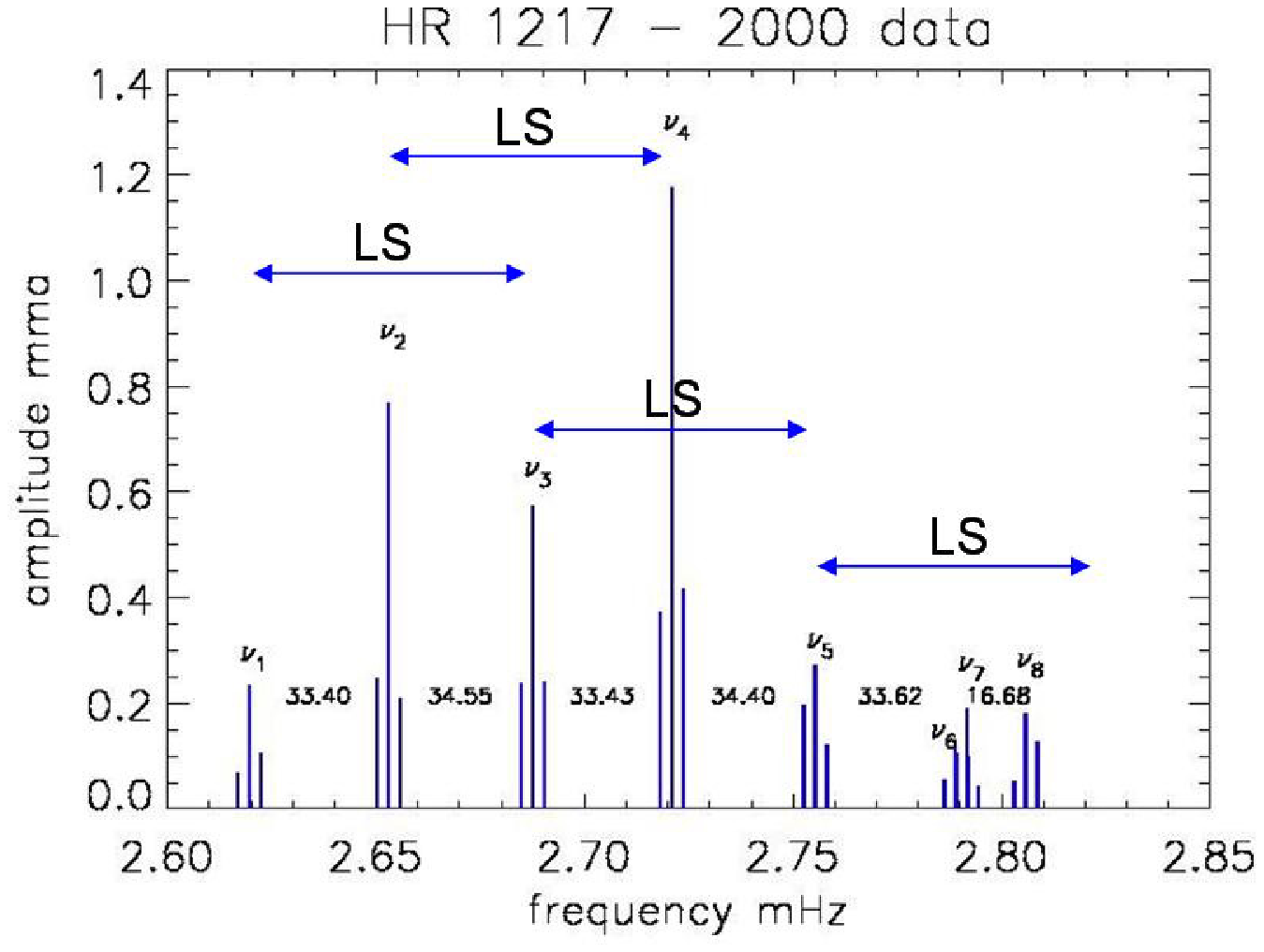}{Schematic oscillation spectra of HR1217, according to the observations obtained during the WET campaign \citep{kurtz05}. The average large separation (LS) derived from the spacing of the first six modes of oscillation is shown by arrows positioned at different frequencies. The strange spacing of the last mode and the asymmetry in the amplitudes of the peaks in the multiplets are clearly seen.}{hr1217}{!ht}{clip,angle=0,width=80mm}

Over the past decade several theoretical works have been carried out with the aim of understanding the effect of the magnetic field on the eigenfrequencies and eigenfunctions of roAp stars \citep{dziembowski96,bigot00,cunha00,saio04,saio05,cunha06}. Two main observational signatures have been given particular attention, namely, the structure of the multiplets and the spacing between consecutive frequencies in the oscillation spectra. Figure \ref{hr1217} shows a schematic view of the oscillation spectra of HR1217. The asymmetry in the amplitude of the peaks in the multiplets and the strange spacing between frequencies $\nu_6$ and $\nu_8$ are evident.

In the case of roAp stars, the multiplet structures seen in the oscillation spectra are due to the modulation of modes of single azimuthal order $m$ over the rotation of the star \citep{kurtz90}. As first shown by \cite{dziembowski96}, the direct effect of the magnetic field modifies the eigenfunctions in such a way that they will no longer be well described by a single spherical harmonic. Consequently, the multiplet structure associated with a mode that in the absence of a magnetic field would be described by a single degree $l$, will in general show additional components, associated with other degrees, which are produced by the magnetic distortion of the eigenfunctions. A few examples of how the multiplets are distorted by the magnetic field when the direct effect of rotation on the oscillations is neglected are given in \cite{saio04}. A comparison between the observed multiplets and the corresponding theoretical expectations can provide information about the inclination between the magnetic field axis and the rotation axis, and between the latter and the line of sight, as well as about the topology of the magnetic field. 

 Another aspect of the observed multiplets structure that has been of concern to theoretical studies is the asymmetry in the multiplet peaks. This asymmetry cannot be explained by the action of the magnetic field on the oscillations, because it can only be produced by an agent, such as the Coriolis force, which can distinguish between the north and south hemisphere \citep[e.g.][]{bigot02,gough05,cunha05}. In this context \cite{bigot02} considered the combined effect of rotation and magnetic field on the oscillations of roAp stars and have shown that if the magnetic and centrifugal effects on the oscillation frequencies are comparable, then the axis of pulsation is no longer aligned with the magnetic axis. Moreover, in this case, even though the Coriolis effect is much smaller than the effects of both the magnetic field and the centrifugal distortion of the star, the former has the important observational consequence of providing a natural explanation for the asymmetry of the peaks in the multiplets. However, in most well studied roAp stars the magnetic field is considerably stronger than that considered by the authors. When the magnetic perturbation dominates over the centrifugal perturbation, the  pulsation axis is expected to be closely aligned with the magnetic axis, and the asymmetry produced by the Coriolis effect disappears. However, it has been known from the works of \cite{cunha00,cunha06} and \cite{saio04} that the magnetic effect on the oscillation frequencies varies in a cyclic way with both magnetic strength and oscillation frequency, alternating between maxima and minima. Thus, even at very strong magnetic fields, if the magnetic effect on the oscillations becomes sufficiently small to be comparable with the perturbation due to centrifugal distortion, it might still be possible to see the effect of the Coriolis force in the structure of the multiplet.

The direct effect of the magnetic field on the frequencies of the oscillations has also been the subject of several studies over the past years. Earlier results have shown that the oscillation frequencies, the large, and the small separations are expected to be significantly modified by the action of the magnetic field \citep{cunha00,saio04}, and that anomalies in the frequency spacing such as that observed in the highest frequency mode of HR1217, might be qualitatively explained by that effect.  Moreover, it became clear that mode conversion in the magnetic boundary layer leads to energy losses that can be relatively large at particular frequencies \citep{cunha00}, and that this dissipation helps explain the absence of $\delta$-scuti type pulsations in roAp stars \citep{saio05}. More recently, \cite{cunha06} has shown that it might be possible to derive, from the observed perturbations, information about the magnetic field properties in the magnetic boundary layer. In particular, this work shows that the oscillation frequencies are influenced by the magnetic field in two distinct ways: firstly the magnetic frequency shifts scale with frequency in a way that depends essentially on the structure of the outer layers and the intensity of the magnetic field; secondly, the amount by which the real part of the frequency shift jumps at well defined frequencies depends essentially on the magnetic field configuration and on the degree of the mode. This separation between the effects of magnetic strength and magnetic topology, diminishes, considerably, the number of models that have to be considered when trying to match the oscillation spectra of a given roAp star.

\subsection*{2.3 Indirect effect of the magnetic field}

Besides its direct effect, the magnetic field can also influence the pulsations in an indirect way, in particular through its interaction with envelope convection. Earlier works \citep{balmforth01,cunha02}, have shown that if convection is suppressed in the envelope of roAp stars, then high frequency oscillations, with periods similar to those observed, could be excited by the opacity mechanism acting on the hydrogen ionization region. Despite this success, the lack of observed $\delta$-Scuti type pulsations in roAp stars, which were also predicted to be excited in the model with convection suppressed, remained unexplained. Recently, it has been shown by \cite{saio05} that the direct effect of the magnetic field  on the oscillations could lead to the stabilization of the low radial order $\delta$-Scuti type pulsations in roAp stars, through the dissipation of slow Alfv\'en waves. Moreover, the effect of diffusion, considered by \cite{theado05,cunha04}, is also expected to help such stabilization. In the absence of envelope convection, the helium settles very quickly in the outer layers, leaving hardly any helium in the region where it undergoes its second ionization. Since the  $\delta$-Scuti type pulsations are excited predominantly by the opacity mechanism acting in the region of second helium ionization, the reduced abundance of helium in that region leads, in most models, to the suppression of this type of pulsation.

\section*{3 Discussion and expectations for the future}

From their discovery, roAp stars have been considered to be particularly well suited for asteroseismic studies, due to the high radial order of their oscillations. While over the past decade theoretical studies have shown that the interpretation of the oscillation spectra of roAp stars is not as straight forward as one could naively have thought, the same studies have revealed the potential of using these observations to learn about the magnetic field of these stars. 

Over the past few years the Canadian satellite MOST has observed four roAp stars, including the well known HR1217 (Matthews, {\it these proceedings}). Through the comparison of these observations and theoretical results obtained with models of roAp stars, we will hopefully improve our understanding of the interaction between the magnetic field and pulsations, and will be able to infer information about the sub-photospheric layers of these stars. Moreover, the launch of the French-led mission CoRoT in December 2006, is expected to bring new insights into studies of roAp stars. As part of its additional science programme, it is hoped that CoRoT will find new roAp stars, which will help establish the observational instability strip for this class of pulsators and test theoretical predictions made through linear stability analysis \citep{cunha02}. Last, but not least, as high resolution spectroscopic observations of roAp stars continue to produce new intriguing results, further theoretical work aimed at understanding the pulsations in their atmospheres is certainly expected.


\acknowledgments{
This work was supported by FCT and FEDER (POCI2010) through the project POCTI/CTE-AST/57610/2004, by FULBRIGHT, through a grant under the Mutual Educational Exchange Program, and by NCAR, through the ECSA and HAO Visiting Scientist Programs.
}


\References{

\bibitem[\protect\citeauthoryear{Balmforth et al.}{2001}]{balmforth01}
Balmforth N.~J. et al., 2001, MNRAS 323, 362 

\bibitem[\protect\citeauthoryear{Balmforth \& Gough}{1990}]{balmforth90}
Balmforth N.~J., Gough D.~O., 1990, ApJ 362, 256 

\bibitem[\protect\citeauthoryear{Bigot et al.}{2000}]{bigot00}
   Bigot, L. et al., 2000, A\&A 356, 218

\bibitem[\protect\citeauthoryear{Bigot \& Dziembowski}{2002}]{bigot02}
   Bigot, L. \& Dziembowski, W. A., 2001, A\&A 391, 235

\bibitem[\protect\citeauthoryear{Cunha \& Gough}{2000}]{cunha00}
   Cunha, M.S. \& Gough, D.O., 2000, MNRAS 319, 1020

\bibitem[\protect\citeauthoryear{Cunha}{2002}]{cunha02}
    Cunha, M.S., 2002, MNRAS 333,47

\bibitem[\protect\citeauthoryear{Cunha}{2003}]{cunha03}
   Cunha, M.S., 2003, in M.J.Thompson and J.Christensen-Dalsgaard (eds), Stellar Astrophysical Fluid Dynamics. Cambridge Univ. Press, UK, p. 51

\bibitem[\protect\citeauthoryear{Cunha}{2005}]{cunha05}
   Cunha, M.S., 2005, JAA 26, 213

\bibitem[\protect\citeauthoryear{Cunha et al.}{2005}]{cunha04}
   Cunha, M.S., Theado, S., Vauclair, S., 2004, in J. Zverko, J. Ziznovsky, S.J. Adelman, and W.W. Weiss eds, The A-Star Puzzle. Cambridge Univ. Press, UK, p.359

\bibitem[\protect\citeauthoryear{Cunha}{2006}]{cunha06}
   Cunha, M.S., 2006, MNRAS 365, 153

\bibitem[\protect\citeauthoryear{Dziembowski \& Goode}{1996}]{dziembowski96}
   Dziembowski, W. A. \&  Goode, P. R., 1996, ApJ {\bf 458}, 33

\bibitem[\protect\citeauthoryear{Elkin et. al}{2005}]{elkin05}
   Elkin, V.G. et al., 2005, MNRAS 358, 665

\bibitem[\protect\citeauthoryear{Gough}{2005}]{gough05}
 Gough, D.O., 2005, in {\it The Roger Tayler Memorial lectures}, Astronomy \& Geophysics, special issue, Sue Boweler (ed), Royal Astronomucal Society (pub.), p.16

\bibitem[\protect\citeauthoryear{Hubrig et al.}{2004}]{hubrig04}
  Hubrig, S., Szeifert, T., Sch\"{o}ller, M., Mathys, G., Kurtz, D.W., 2004, A\&A 415, 685

\bibitem[\protect\citeauthoryear{Hubrig et al.}{2005}]{hubrig05}
   Hubrig, S. et al., 2005, A\&A 440, L37

\bibitem[\protect\citeauthoryear{Kochukhov}{2006}]{kochukhov06}
Kochukhov, O., 2006, A\&A 454, 321

\bibitem[\protect\citeauthoryear{Kurtz}{1982}]{kurtz82}
Kurtz D.~W., 1982, MNRAS 200, 807

\bibitem[\protect\citeauthoryear{Kurtz}{1990}]{kurtz90}
Kurtz D.~W., 1990, ARA\&A 28, 607

\bibitem[\protect\citeauthoryear{Kurtz \& Martinez}{2000}]{kurtz00}
Kurtz D. W., Martinez P., 2000, BaltA, 9, 253

\bibitem[\protect\citeauthoryear{Kurtz et al.}{2004}]{kurtz04}
Kurtz, D.~W. et al., 2004, in Zverko J., Ziznovsky J., Adelman S. J., Weiss W. W., eds, The A-Star Puzzle. Cambridge Univ. Press, UK, p. 3432

\bibitem[\protect\citeauthoryear{Kurtz et al.}{2005}]{kurtz05}
Kurtz D.~W. et al., 2005, MNRAS 358, 651

\bibitem[\protect\citeauthoryear{Mathys \& Hubrig}{1997}]{mathys97}
  Mathys G. \& Hubrig S., 1997, A\&AS, 124, 475

\bibitem[\protect\citeauthoryear{Ryabchikova et al.}{2006}]{ryabchikova06}
Ryabchikova, T. et al., 2006, A\&A, 445, L47 

\bibitem[\protect\citeauthoryear{Saio \& Gautschy}{2004}]{saio04}
  Saio, H. \& Gautschy, A., 2004, MNRAS 350, 485

\bibitem[\protect\citeauthoryear{Saio}{2005}]{saio05}
   Saio, H., 2005, MNRAS 360, 1022

\bibitem[\protect\citeauthoryear{Theado et al.}{2005}]{theado05}
Theado, S., Vauclair, S., Cunha, M.S., 2005, A\&A 443, 627

}



\end{document}